\documentclass[aps,reprint,prappl,groupedaddress,superscriptaddress,showpacs,longbibliography]{revtex4-1}

\usepackage{natbib}
\usepackage[latin1]{inputenc}
\usepackage{graphicx}
\usepackage{amsmath,amssymb,stmaryrd,bm,dcolumn,textcomp}
\usepackage{mathrsfs}
\usepackage{rotating,array,color,float,braket}
\usepackage{xcolor}
\usepackage{hyperref}					
\hypersetup{							    
    colorlinks,
    linkcolor={blue!80!black},
    citecolor={blue!80!black},
    urlcolor={blue!80!black}
}

\begin{document}

\title{Acoustic Resonance and Streaming\\
in Droplet Acoustofluidics}

\author{Fabio Garofalo}
\email[corresponding author:]{fabio.garofalo@bme.lth.se}
\affiliation{Department of Biomedical Engineering, Lund University,
Ole R\"omers V\"ag 3 S-22363, Lund, Sweden}

\begin{abstract}
This paper reports the phenomenon of resonance weakening and streaming onset in two phase acoustofluidics by performing numerical simulations of a capillary droplet suspended in a microfluidic chamber.
The simulations show that depending on the relative acoustic properties of the two phases, it is possible to observe (i)~the decrease in the total acoustic energy as the oscillation amplitude at the wall increases, and (ii)~the onset of acoustic streaming.
The impact of these findings in terms of acoustic focusing inside droplets is also discussed.
\end{abstract}


\maketitle

\section{Introduction}
Acoustofluidics is an established microfluidics technology that combining ultrasounds and a carrier flow allows for a fine manipulation of particle at microscale \cite{Evander_2012,Augustsson_2012,Nordin_2012,Manneberg_2008,Augustsson_2012_Anal_Chem, Ding_2014, Petersson_2007}.

Recently, some researchers attempted to combine acoustofluidics and droplet acoustofluidics for particle focusing inside droplets.
The results for standard fluids used in droplet microfluidics, i.e. fluorinated oil HFE-7500, have been subjected of controversies. Indeed, it seems that other organic phases have shown better performance for particle focusing inside the droplets.

The present paper employs numerical simulations to give a first insight on the phenomena involved in acoustic resonance and streaming in droplet acoustofluidics. Specifically, simulations of a water droplet in HFE-7500 and olive oil have been performed.

\section{Model Setting}

\begin{figure}[!!h]
\center
\begin{picture}(233,170)
\put(0,0){
\includegraphics[width=8cm]{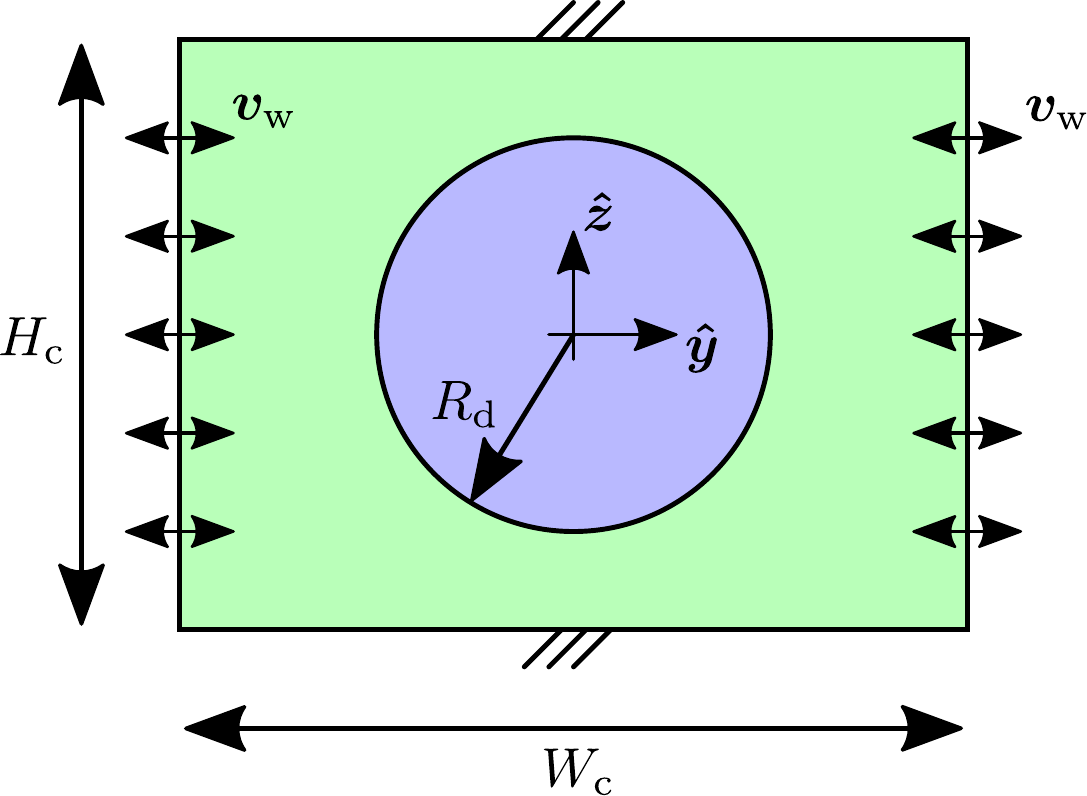}}
\end{picture}
\caption{Geometry of the two-dimensional microfluidic chamber used in the simulations.}
\label{fig:schematic}
\end{figure}

The geometry used in the simulation is that reported in Fig.~\ref{fig:schematic}. A water droplet (suffix ``'w' means water properties) is suspended in an organic phase (suffixes ``OO'' and ``HFE'' refer to olive oil and HFE respectively) within a microfluidic chamber. The vertical channel walls oscillate with a predetermined displacement amplitude.

\paragraph*{Oscillation Amplitude at Chamber Walls.}
Since the	piezo and the material embedding the microchannel are neither modeled nor their response is simulated, there is the need to assume the acoustic pressure amplitude at the resonance frequency. This implies also that there is no feedback to the transducer, namely no change in the overall impedance.
The driving pressure rescaling is performed by computing the resonance acoustic frequency and then rescaling the driving amplitude at the boundary proportionally to the desired value of the acoustic pressure amplitude $p_\mathrm{ac}$, defined as the maximum of the acoustic pressure. The hypothetical acoustic pressure amplitudes considered in the simulations are in the range \mbox{$p_\mathrm{ac}^{}=0.01-5\,\mathrm{MPa}$}.

\subsection{Governing Equations}
The governing equations are the mass density conservation and the linear momentum density balance,
\begin{subequations}\label{eq:gov}
\begin{align}
\partial_t^{}\rho_{}^{}&=\,-\bm{\nabla}_{}^{}\cdot\left(\rho_{}^{}\bm{v}_{}^{}\right)\,,\\
\partial_t^{}\left(\rho_{}^{}\bm{v}_{}^{}\right)&=\,-\bm{\nabla}_{}^{}\cdot\bm{\Pi}_{}^{}+\bm{f}_{\mathrm{v}}^{}\,,
\end{align}
\end{subequations}
where $\rho_{}^{}$ is the fluid density, $\bm{v}_{}^{}$ is the fluid velocity, $\bm{f}_\mathrm{v}^{}$ is a volume force (there is a plan to include gravity, computing the initial configuration for the droplet and then run the time-dependent simulations), and
\begin{equation}
\bm{\Pi}_{}^{}=\rho_{}^{}\bm{v}_{}^{}\bm{v}_{}^{}-\bm{\sigma}_{}^{}\,,
\end{equation}
is the momentum (density) flux. In the latter relation, \mbox{$\rho_{}^{}\bm{v}_{}^{}\bm{v}_{}^{}$} is the advective momentum flux while $\bm{\sigma}_{}^{}$ is the stress tensor
\begin{equation}
\bm{\sigma}_{}^{}=-\left(p_{}^{}-\lambda_{}^{}\bm{\nabla}_{}^{}\cdot\bm{v}_{}^{}\right)\,\bm{I}_{}^{}+\eta_{}^{}\left[\bm{\nabla}_{}^{}\bm{v}_{}^{}+\left(\bm{\nabla}_{}^{}\bm{v}_{}^{}\right)^T\right]\,,
\end{equation}
in which $p_{}^{}$ is the pressure, $\eta_{}^{}$ is the shear viscosity, and $\lambda_{}^{}$ is the dilational viscosity (values for this parameter are difficult to find, we consider either \mbox{$\lambda_{}^{}=0$} or by Stokes' approximation \mbox{$\lambda\simeq\tfrac{2}{3}\eta_{}^{}$}).

Equations need to be equipped with the boundary conditions. Here a rectangular closed domain (see Fig.~\ref{fig:schematic}) with two vertical walls at \mbox{$y=\pm W_c^{}/2$} is considerd. The vertical walls oscillate at angular frequency $\omega_{}^{}$, and thus it can write for the wall velocity $\bm{v}_\mathrm{w}^{}$
\begin{equation}
\bm{v}_\mathrm{w}^{}(\pm W_c/2,z)=[U(z), 0]^T\,e^{-i\,\omega\,t}\,,
\end{equation}
where $U_{}^{}(z)$ is chosen such that the simulation yields the desired pressure amplitude, that is $p_\mathrm{ac}^{}$. At the bottom and top walls the rigid wall conditions
\begin{equation}
\bm{v}(y,\pm H_c/2)=\bm{0}_{}^{}\,,
\end{equation}
are assigned.
\subsection{Interface Conditions}
At the interface between the continous phase and disperse phase (the droplet) the interface conditions
\begin{subequations}
\begin{align}
\bm{0}&=\,-\llbracket\bm{v}_{}^{}\rrbracket\,,\\
\bm{0}&=\,-\llbracket\bm{\Pi}_{}^{}\rrbracket\cdot\bm{\hat{n}}_{}^{}+\bm{f}_{\mathrm{st}}^{}\,,
\end{align}
\end{subequations}
are considered ($\bm{\hat{n}}_{}^{}$ is the outward-pointing normal). The surface tension force is given by
\begin{equation}
\bm{f}_{\mathrm{st}}^{}=\gamma_{}^{}\left(\bm{\nabla}_{\mathrm{s}}^{}\cdot\bm{\hat{n}}_{}^{}\right)\bm{\hat{n}}_{}^{}\,.
\end{equation}
The symbol $\bm{\nabla}_{\mathrm{s}}^{}$ denotes the surface nabla operator,
\begin{equation}
\bm{\nabla}_{\mathrm{s}}^{}=\left(\bm{I}_{}^{}-\bm{\hat{n}}_{}^{}\bm{\hat{n}}_{}^{}\right)\cdot\bm{\nabla}_{}^{}\,,
\end{equation}
where $\bm{I}_{}^{}$ is the unit tensor ($[\bm{I}_{}^{}]_{ij}=\delta_{ij}$).

\paragraph*{Weak form for Surface Tension at the Interface.}
In the presence of surface tension force and acoustic force, the weak form contribution at the interface is given by
\begin{equation}
\mathscr{L}_{\mathrm{hd}}^{\mathrm{int}}=-\gamma_{}^{}\,\mathrm{Tr}_{}^{}\left[\left(\bm{I}_{}^{}-\bm{\hat{n}}_{}^{}\bm{\hat{n}}_{}^{}\right)\cdot\bm{\nabla}_{}^{}\bm{v}_0^{}
\right]+\bm{f}^{}_\mathrm{ac}\cdot\bm{v}_0\,,
\end{equation}
which is true just for a closed interface. Note that there is no need to compute the interface curvature since in the weak form this will not appear.

\subsection{Time-scale Splitting}
It is assumed that the hydrodynamic time-scale and the acoustic time-scale are well separated and correspond to a slow time-scale and a fast time-scale, respectively. The slow-time scale dynamics is described by means of the incompressibility constraint and the linear momentum density balance for the time-average density $\rho_0^{}$ and velocity $\bm{v}_0^{}$,
\begin{subequations}
\begin{align}
0_{}^{}&=\,-\bm{\nabla}_{}^{}\cdot\bm{v}_{0}^{}\,,\\
\partial_t^{}\left(\rho_0^{}\bm{v}_0^{}\right)&=\,-\bm{\nabla}_{}^{}\cdot\bm{\Pi}_{0}^{}+\bm{f}_{\mathrm{v}0}^{}\,.
\end{align}
\end{subequations}
The time-averaged momentum flux $\bm{\Pi}_0^{}$ is
\begin{equation}
\bm{\Pi}_0^{}=\rho_0\bm{v}_0^{}\bm{v}_0^{}-\bm{\sigma}_0^{}\,,
\end{equation}
where
\begin{equation}
\bm{\sigma}_0^{}=-p_0^{}\bm{I}+\eta_{0}^{}\left[\bm{\nabla}_{}^{}\bm{v}_0+\left(\bm{\nabla}_{}^{}\bm{v}_0^{}\right)^T\right]\,.
\end{equation}
For the slow time-scale we consider the no-slip conditions at the channel walls
\begin{equation}
\bm{v}_0^{}=\bm{0}_{}^{}\,.
\end{equation}
The slow time-scale interface conditions are
\begin{subequations}
\begin{align}
\bm{0}&=\,-\llbracket\bm{v}_0^{}\rrbracket\,,\\
\bm{0}&=\,-\llbracket\bm{\Pi}_0^{}\rrbracket\cdot\bm{\hat{n}}_{}^{}+\bm{f}_{\mathrm{ac}}+\bm{f}_{\mathrm{st}}^{}\,,
\end{align}
\end{subequations}
where $\bm{f}_{\mathrm{ac}}^{}$ is the acoustic force at the interface
\begin{equation}
\bm{f}_{\mathrm{ac}}^{}=-\llbracket\braket{\bm{\Pi}_2^{}}\rrbracket\cdot\bm{\hat{n}}_{}^{}\,.
\end{equation}
The time-average of the second-order momentum density flux is given by
\begin{equation}
\braket{\bm{\Pi}_2^{}}=
\frac{\braket{\bm{\nabla}_{}^{}p_1^{}\bm{\nabla}_{}^{}p_1^{}}}{\rho_0^{}\omega_{}^2}
+\left(\frac{\braket{p_1^2}}{2\,\rho_0^{}c_0^2}-
\frac{\braket{|\bm{\nabla}_{}^{}p_1^{}|_{}^2}}{2\,\rho_0^{}\omega_{}^2}
\right)\bm{I}
\end{equation}
or in terms of complex-valued pressure amplitude $\hat{p}_1^{}$
\begin{equation}
\braket{\bm{\Pi}_2^{}}=
\frac{\bm{\nabla}_{}^{}\hat{p}_1^{}\bm{\nabla}_{}^{}\hat{p}_1^{}}{2_{}^{}\rho_0^{}\omega_{}^2}
+\left(\frac{\hat{p}_1^*\hat{p}_1^{}}{4_{}^{}\rho_0^{}c_0^2}-
\frac{\bm{\nabla}_{}^{}\hat{p}_1^*\cdot\bm{\nabla}_{}^{}\hat{p}_1^{}}{4_{}^{}\rho_0^{}\omega_{}^2}
\right)\bm{I}
\end{equation}
where the amplitude of the first-order acoustic pressure, i.e. $p_1^{}(\bm{x},t)=\hat{p}_1^{}(\bm{x},\omega)\,e^{-i\,\omega\,t}$, is the solution of the Helmholtz equation
\begin{equation}
\frac{\hat{p}_1^{}}{\rho_0^{}c_0^2}+\bm{\nabla}_{}^{}\cdot\left(\frac{\bm{\nabla}_{}^{}\hat{p}_1^{}}{\rho_0^{}\omega_{}^2}\right)=0\,,
\end{equation}
with the boundary condition
\begin{equation}
\bm{\nabla}_{}^{}\hat{p}_1^{}\cdot\bm{\hat{n}}_{}^{}=\mathrm{i}_{}^{}\rho_0^{}\omega[U(z), 0]^T\,,
\end{equation}
at the vertical walls and
\begin{equation}
\bm{\nabla}_{}^{}\hat{p}_1^{}\cdot\bm{\hat{n}}_{}^{}=0\,,
\end{equation}
at the bottom and top walls. The fast-time scale interface conditions are
\begin{subequations}
\begin{align}
\llbracket \hat{p}_1^{} \rrbracket&=\,0\,,\\
\llbracket \rho_0^{-1}\bm{\nabla}_{}^{}\hat{p}_1^{} \rrbracket\cdot\bm{\hat{n}}_{}^{}&=\,0\,,
\end{align}
\end{subequations}

\section{Results}
Table \ref{tab:physprop} reports the physical properties used in the num erical simulations.
\begin{table}[!!t]
\caption{List of the physical properties used in the numerical simulations.}
\begin{tabular}{lcccc}
\label{tab:physprop}
 & $\rho_{0}^{}\,[\mathrm{kg_{}^{}/m_{}^3}]$ & $c_{0}^{}\,[\mathrm{m_{}^{}/s_{}^{}}]$ & $\eta_{0}^{}\,[\mathrm{mPa_{}^{}\cdot s_{}^{}}]$ & $\gamma_{\mathrm{W},x}^{}\,[\mathrm{mN_{}^{}/m_{}^{}}]$ \\
Water & $997$ & $1497$ & $1.0$ & $-$ \\
Olive Oil & $910$ & $1450$ & $61.0$ & $16.0$ \\
HFE-7500 & $1619$ & $660$ & $1.2$ & $41.0$ \\
Silicon Oil & $950$ & $1100$ & $19.0$ & $32.0$
\end{tabular}
\end{table}
\begin{figure}[!!h]
\center
\begin{picture}(233,298)
\put(-55,-20){
\includegraphics[width=10cm]{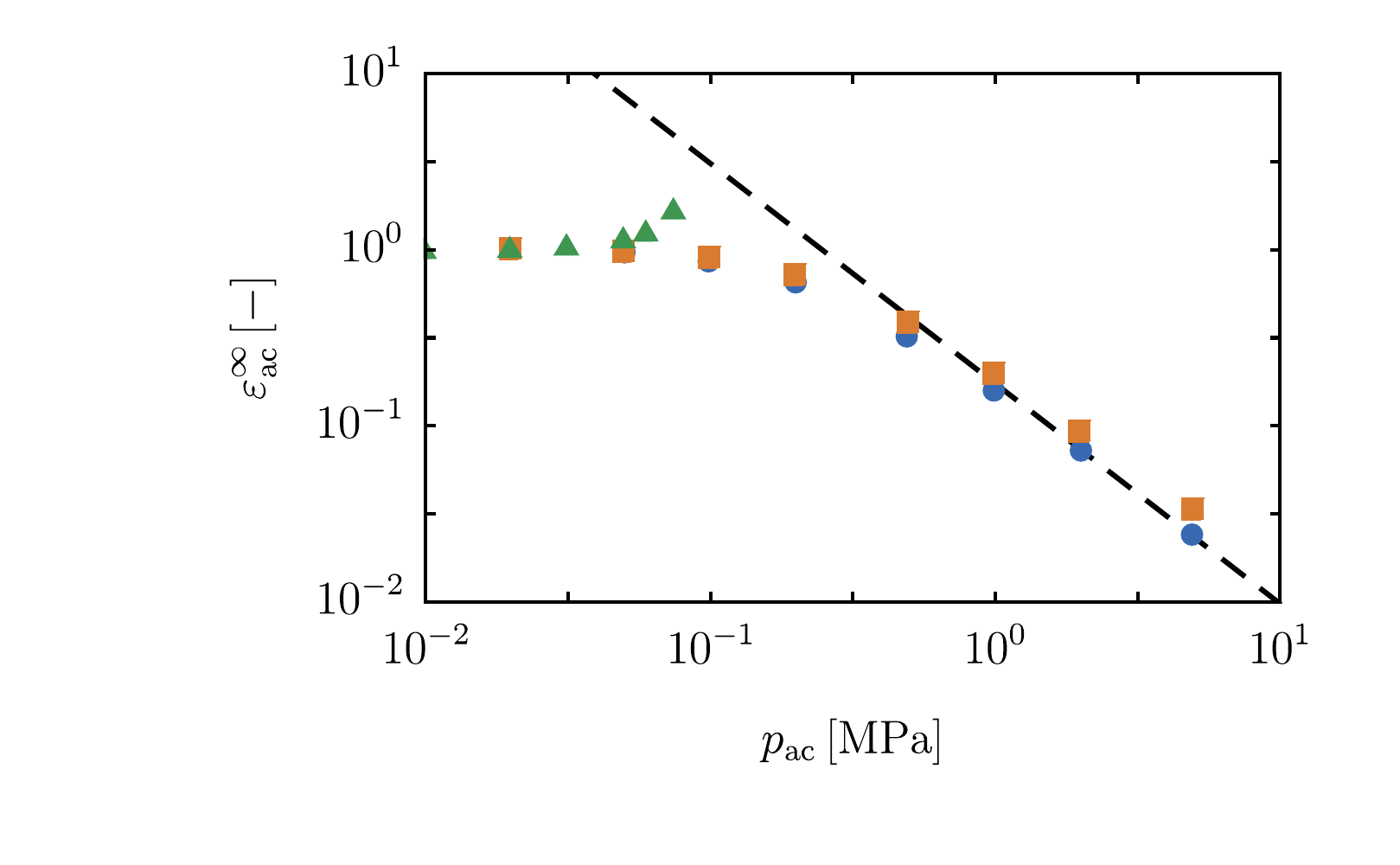}}
\put(-55,135){
\includegraphics[width=10cm]{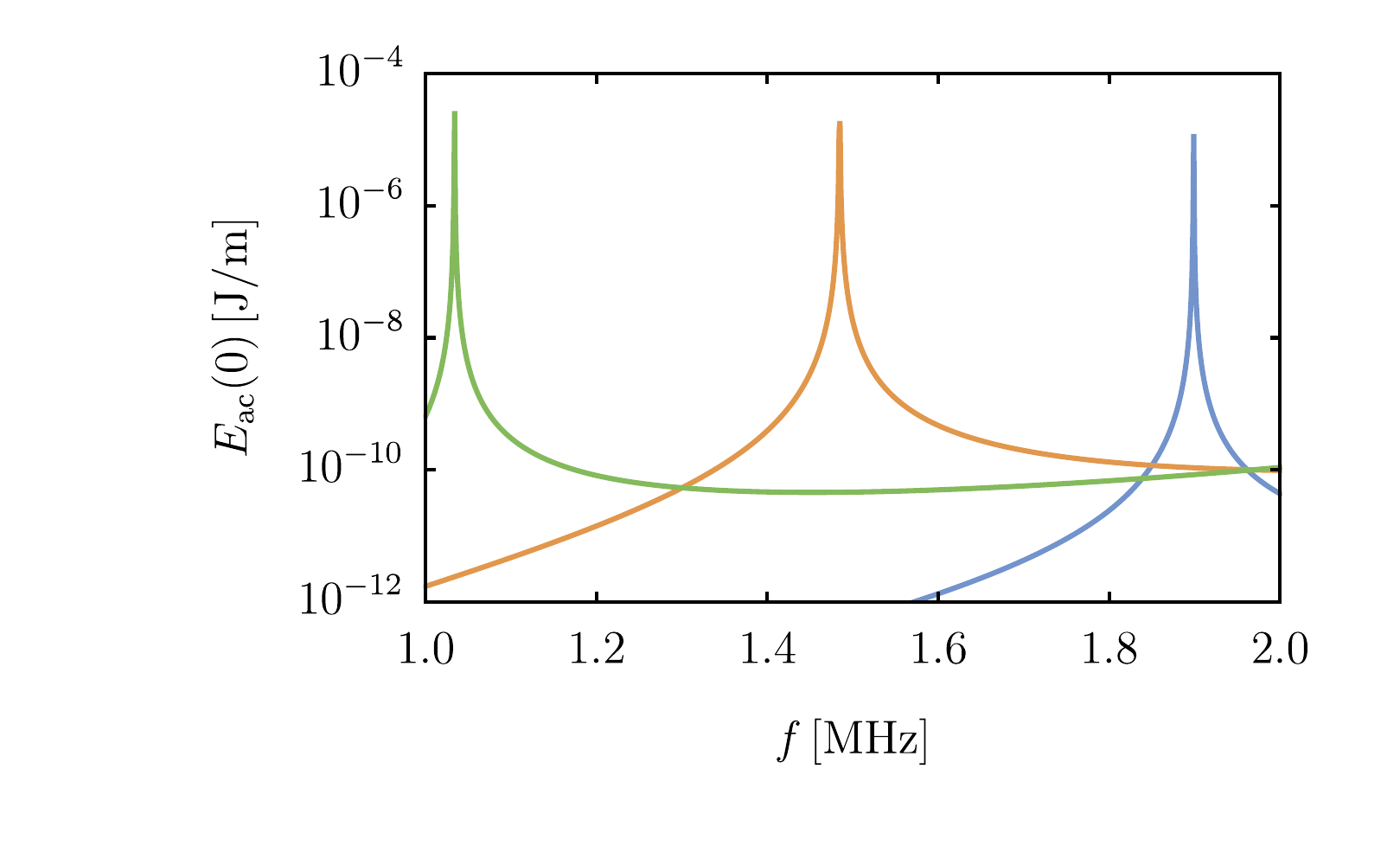}}
\put(194,285){(a)}
\put(194,130){(b)}
\end{picture}
\caption{(color online) (a) Initial Acoustic Energy $E_\mathrm{ac}(0)$ with $p_\mathrm{ac}=1\,\mathrm{MPa}$ as function of the driving frequency $f$ for the three systems: (blue) OO/Water, (orange) SO/Water, and (green) HFE/Water. (b) Final to initial acoustic energy ratio $\varepsilon_\mathrm{ac}^{}$ as function of the driving acoustic pressure amplitude $p_\mathrm{ac}^{}$: \mbox{OO/Water ({\LARGE{\raisebox{-0.2ex}{$\bullet$}}})}, \mbox{SO/Water (\raisebox{-.2ex}{$\blacksquare$})}, and \mbox{HFE/Water ($\blacktriangle$)}. Black dashed line is a reference slope $\varepsilon_\mathrm{ac}^\infty\sim p_\mathrm{ac}^{-5/4}$.}
\label{fig:resweak}
\end{figure}

Figure \ref{fig:resweak} shows the resonance spectrum (a) in the case of a water droplet suspended in olive oil, silicone oil, and HFE oil (see caption for color coding), by plotting the acoustic energy $E_\mathrm{ac}(0)$ as function of the frequency $f$.
The three resonance frequencies set in the beginning of the time-dependent simulations were (put the resonance frequencies here).

Figure \ref{fig:resweak} shows the ratio between the acoustic energy when the droplet has reach the final stage and the initial acoustic energy, that is $\varepsilon^\infty_\mathrm{ac}$, as function of the rescaled driving acoustic pressure at the walls $p_\mathrm{ac}$.
It has been found that as the driving pressure increases the final acoustic energy is less than that expected for a immutable geometric configuration, in the case of olive oil and silicone oil.
In the case of HFE the acoustic energy has shown a small increase before to reach a value for the driving pressure $p_\mathrm{ac}$ beyond that the droplet dynamics is unstable.

In the case of the stable simulations for HFE/water system, it has been found that there is an onset of acoustic streaming.
\begin{figure}[!!t]
\center
\begin{picture}(233,298)
\put(-55,-20){
\includegraphics[width=10cm]{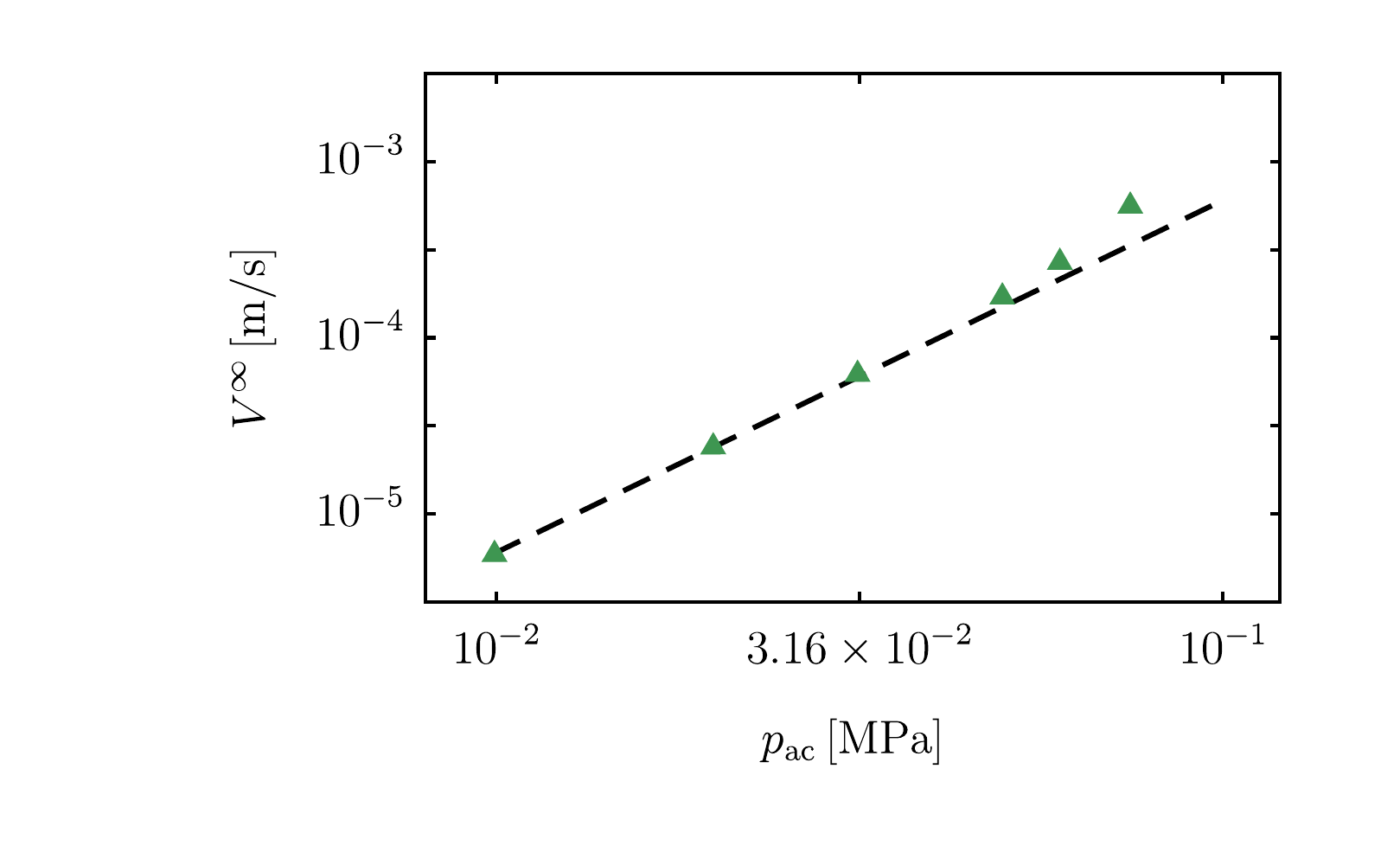}}
\put(-55,135){
\includegraphics[width=10cm]{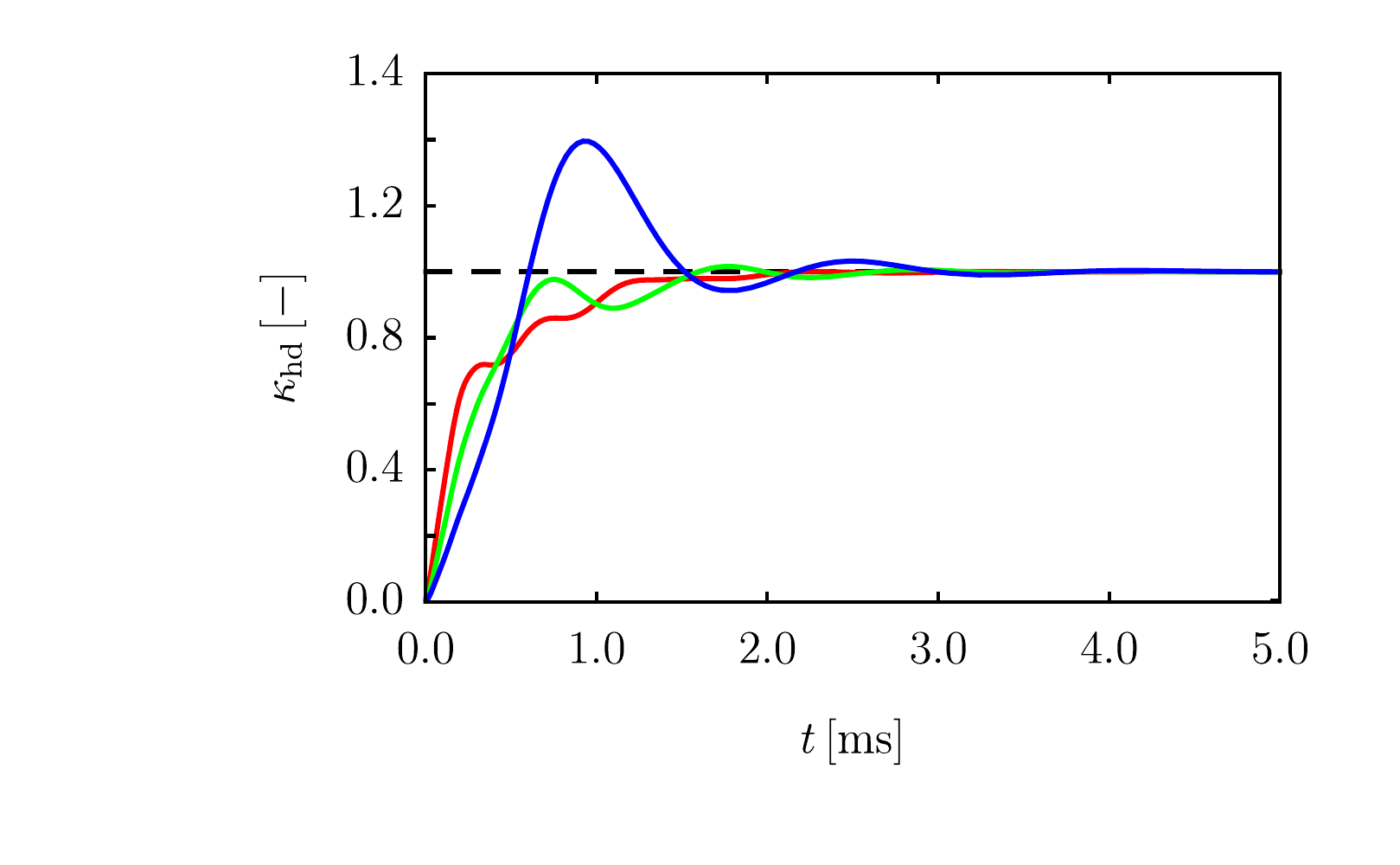}}
\put(194,285){(a)}
\put(194,130){(b)}
\end{picture}
\caption{(color online) (a) Hydrodynamic kinetic energy ratio $\kappa_\mathrm{hd}^{}$ as function of time $t$ for HFE/Water system at three driving pressure amplitudes: (red) $p_\mathrm{ac}^{}=10^{-2}\,\mathrm{MPa}$, (green) $p_\mathrm{ac}^{}=6\times 10^{-2}\,\mathrm{MPa}$, and (blue) $p_\mathrm{ac}^{}=7.5\times 10^{-2}\,\mathrm{MPa}$. (b) Asymptotic average velocity $V_{}^\infty$ for HFE/Water system as function of the driving pressure amplitude $p_\mathrm{ac}^{}$. Black dashed line is a reference slope $V^\infty\sim p_\mathrm{ac}^{2}$.}
\label{fig:hfestream}
\end{figure}
The evidence of this is reported in figure \ref{fig:hfestream} that shows the ratio between the kinetic energy and the final kinetic energy as function of the time (a), together with the average fluid velocity at the final time.

The phenomenon of acoustic streaming onset has not been observed for OO/water and SO/water system.
\begin{figure}[!!h]
\center
\begin{picture}(233,298)
\put(-55,-20){
\includegraphics[width=10cm]{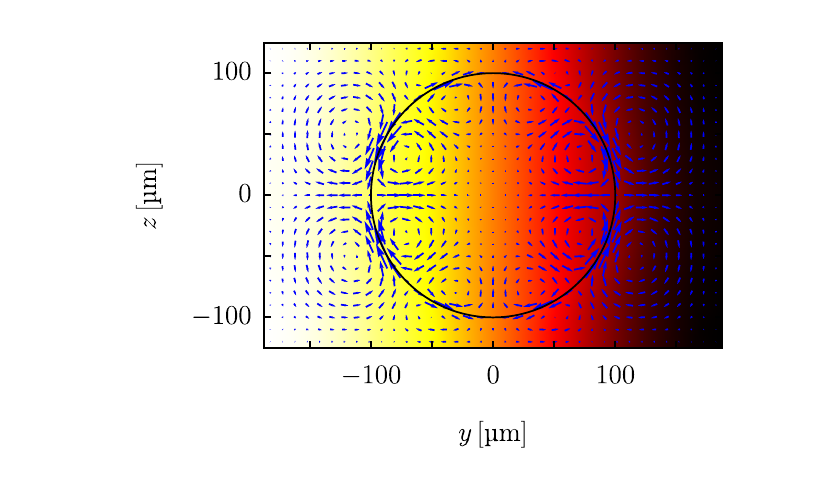}}
\put(-55,135){
\includegraphics[width=10cm]{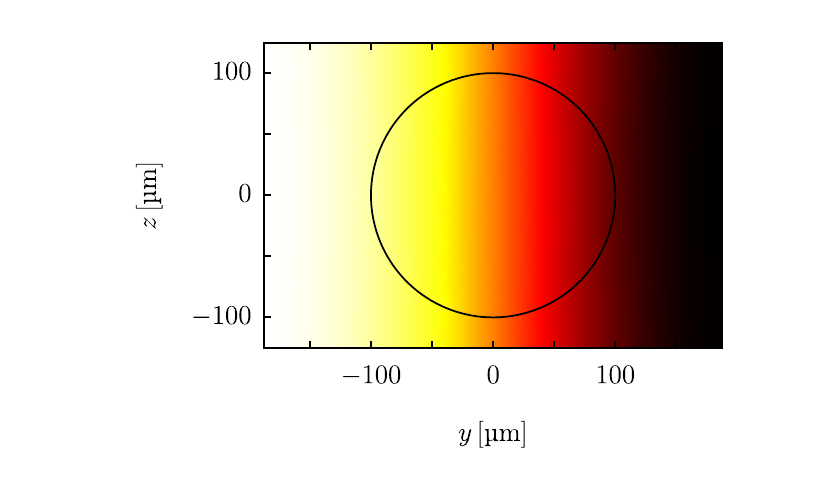}}
\put(187,285){\begin{color}{white}(a)\end{color}}
\put(187,130){\begin{color}{white}(b)\end{color}}
\end{picture}
\caption{(color online) Acoustic pressure field $\hat{p}_1$ (color shade) and velocity field $\bm{v}_0^{}$ (blue arrows) at the asymptotic for OO/Water system at $p_\mathrm{ac}^{}=0.1\,\mathrm{MPa}$ (a) and HFE/Water system at $p_\mathrm{ac}^{}=0.075\,\mathrm{MPa}$. The acoustic pressure field is rescaled with the maximum value.}
\label{fig:cfdplot}
\end{figure}
The evidence of this fact has been reported in figure \ref{fig:cfdplot} where the together with the rescaled pressure field (color shading), also the normalized acoustic streaming velocity field is reported (blue arrows).

\section{Discussion}
As one studies the behavior of the acoustic energy for a closed system where a droplet of water is immersed in an organic phase, two different different phenomenologies can be observed.
These depend on on the relative acoustic properties of the two phases.

\subsection{Acoustic Energy}
For OO/water systems, as the driving acoustic pressure increases the energy appears to be initially constant as the interface tension force prevents the particle deformation.
However, when the acoustic force at the interface starts to increase and competing with the interface tension force the droplet is deformed.
As this deformation appears, the driving frequency does not match the resonance frequency of the new geometric configuration and the acoustic energy descreases.

For HFE/water systems, the observations suggest that as the driving pressure amplitude increases, the energy increases until an unstable state (most likely corresponding to a new resonance frequency for the deformed system) is reached. The simulations could not be performed beyond a certain value of the acoustic driving pressure.

\subsection{Acoustic Streaming}
For OO/Water systems, the relative physical properties of the two fluids have shown a suppression of the acoustic streaming.

For HFE/Water systems, it has been observed that there is an onset of the acoustic streaming. This fact deserve more investigations.

\section{Conclusions}
The impact of having a capillary droplet enclosed in a microfluidic and acoustically driven microchamber has been studied.

Two behaviors has been observed and these depend on the relative acoustic properties of the disperse phase (the droplet) and the continous phase (the surrounding fluid).

In the case of a water droplet suspended in olive oil, it has been observed both a weakening of the resonance and streaming suppression.
In the case of a water droplet suspended in HFE oil, it has been observed an increase in the acoustic energy and the onset of streaming.

Further investigations are considered since it is important to provide a meaningful explanation of the observed behaviors and how these impact on particle/cell manipulation inside droplets.

\appendix

\section{Indicators}

(short scale time-averaged)Acoustic Lagrangian Density
\begin{equation}
\mathscr{L}_\mathrm{ac}^{}(\bm{x}_{}^{},t_{}^{})=
\frac{\bm{\nabla}\hat{p}_1^*\cdot\bm{\nabla}\hat{p}_1^{}}{4\,\rho_0^{}\omega_{}^2}
-\frac{\hat{p}_1^*\hat{p}_1^{}}{4\,\rho_0^{} c_0^2}\,,
\end{equation}
Acoustic Power-Input Density
\begin{equation}
\mathscr{P}_\mathrm{ac}=\tfrac{1}{2}\hat{p}_1^*\hat{\bm{v}}_1^{}\cdot\bm{\hat{n}}_{}^{}\,,
\end{equation}
Acoustic Energy Density (as indicator of acoustophoresis)
\begin{equation}
\mathscr{E}_\mathrm{ac}^{}(\bm{x}_{}^{},t_{}^{})=
\frac{\bm{\nabla}\hat{p}_1^*\cdot\bm{\nabla}\hat{p}_1^{}}{4_{}^{}\rho_0^{}\omega_{}^2}+
\frac{\hat{p}_1^*\hat{p}_1^{}}{4_{}^{}\rho_0^{} c_0^2}\,,
\end{equation}
the total acoustic energy is then
\begin{equation}
E_\mathrm{ac}(t)=\int_V \mathscr{E}_\mathrm{ac}^{}(\bm{x}_{}^{},t_{}^{})\,\mathrm{d}V\,,
\end{equation}
as the initial configuration is that used to compute the resonance frequency we can define the ratio between the final acoustic energy and the initial acoustic energy, that depends on $p_\mathrm{ac}^{}$
\begin{equation}
\varepsilon_\mathrm{ac}^{}(t)=\frac{E_\mathrm{ac}^{}(t)}{E_\mathrm{ac}^{}(0)}\,,
\end{equation}
note that for an ``immutable'' geometric configuration this ratio must equals $1$ for any $t$. We define $\varepsilon_\mathrm{ac}^\infty=E_\mathrm{ac}^{}(\infty)/E_\mathrm{ac}^{}(0)$.

\subsection{Indicators}
\subsubsection{Assessment of good simulations}
\begin{itemize}
\item simulation of small droplets dislocation towards to/far from the pressure node (depending on the contrast factor)
\item comparison with \textit{standard} theory of particle motion in acoustic field (give a try to the shift in resonance when the particle/droplet undergoes focusing, it can be interested to show this)
\end{itemize}
\paragraph*{Analysis of Droplet Motion}
Droplet Volume (in 2D this measures an area)
\begin{equation}
V_\mathrm{droplet}^{}=\int_{\mathrm{droplet}}^{}\,\mathrm{d}V_{}^{}
\end{equation}
this in reliable simulations must be a constant, or at least we admit $\pm 5\%$ in variation (or less depending on the computational efforts).
The averaged droplet position can be computed by considering the integral
\begin{equation}
\bm{x}_\mathrm{droplet}^{}=\frac{1}{V_\mathrm{droplet}^{}}\int_\mathrm{droplet}^{}\bm{x}_{}^{}\,\mathrm{d}V_{}^{}
\end{equation}
when we validate the simulations, we compare this quantity against the \textit{theoretical} droplet trajectory given by integration of the equation of motion for a spherical particle subjected to acoustic radiation forces. In the case of the centered droplet we admit a variation pf $\pm 5\%$ (maybe less) from the centered position.
Acoustic force on the droplet
\begin{equation}
\bm{F}_\mathrm{ac}=\int_{S_\mathrm{droplet}}\bm{f}_\mathrm{ac}\,\mathrm{d}S
\end{equation}
to be compared with the known expressions in the case of ``particle-like'' droplet.
\subsubsection{Phenomenon Analysis}
\subsubsection{Acoustics}

\subsubsection{Hydrodynamics}
Hydrodynamic Lagrangian Density
\begin{equation}
\mathscr{L}_\mathrm{hd}=\bm{v}_0^{}\cdot\partial_t^{}(\rho_0^{}\bm{v}_0^{})
-\bm{\Pi}_0^{}:\bm{\nabla}_{}^{}\bm{v}_{0}^{}
\end{equation}
rate of change in the hydrodynamic kinetic energy density
\begin{equation}
\partial_t\mathscr{K}_\mathrm{hd}^{}=\bm{v}_0^{}\cdot\partial_t^{}(\rho_0^{}\bm{v}_0^{})\,.
\end{equation}
Hydrodynamic kinetic energy density,
\begin{equation}
\mathscr{K}_\mathrm{hd}^{}=\frac{1_{}^{}}{2_{}^{}}\rho_0^{}v_0^2\,,
\end{equation}
needed to evaluate the presence/absence of a velocity field in the model.\\
(Rate of) viscous dissipation density
\begin{equation}
\mathscr{D}_\mathrm{hd}(\bm{x}_{}^{},t)=\bm{\Pi}_0^{}:\bm{\nabla}_{}^{}\bm{v}_{0}^{}\,,
\end{equation}
Total kinetic energy density
\begin{equation}
K_\mathrm{hd}(t)=\int_V \mathscr{K}_\mathrm{ac}^{}(\bm{x}_{}^{},t_{}^{})\,\mathrm{d}V\,,
\end{equation}
\begin{equation}
\kappa_\mathrm{hd}(t)=\frac{K_\mathrm{hd}^{}(t)}{K_\mathrm{hd}^{}(\infty)}
\end{equation}
characteristic system velocity
\begin{equation}
V(t)=\sqrt{\frac{2_{}^{}K_\mathrm{hd}(t)}{m}}\,,
\end{equation}
where $m=\int_V\rho(\bm{x},t)\mathrm{d}V=\mathrm{const}$. We also define $V_{}^\infty=V(t\rightarrow\infty)$.
Hydrodynamic power input density (at the interface between the two fluids)
\begin{equation}
\mathscr{P}_{hd}=-\mathscr{L}_\mathrm{hd}^\mathrm{int}
\end{equation}

\subsection*{Lagrangian density for the time-dependent model}
Note that
\begin{equation}
L_{}^{}=\int_0^t\int_V\mathscr{L}_{}^{}(t_{}^{},\bm{x}_{}^{})\mathrm{d}V\,\mathrm{d}t
\end{equation}









\bibliography{acoustofluidics2}

\end{document}